\documentclass [10pt]{article}
\usepackage[left=2cm,top=2.50cm,right=2cm,bottom=2.50cm]{geometry}
\usepackage{mathrsfs}
\usepackage{amsmath,amssymb,latexsym,color,cancel,graphicx,bbm,colortbl}
\usepackage[english]{babel}
\usepackage[latin1]{inputenc}
\usepackage{ragged2e}
\usepackage{cite}

\begin{document}
\date{}

\title{$Sp(4,R)$ algebraic approach of the most general Hamiltonian of a two-level system in two-dimensional
geometry}
\author{E. Chore\~no$^{a}$, D. Ojeda-Guill\'en$^{a,b}$ \footnote{{\it E-mail address:} dojedag@ipn.mx}}
\maketitle

\begin{minipage}{0.9\textwidth}
\small $^{a}$ Escuela Superior de F{\'i}sica y Matem\'aticas,
Instituto Polit\'ecnico Nacional, Ed. 9, Unidad Profesional Adolfo L\'opez Mateos, Delegaci\'on Gustavo A. Madero, C.P. 07738, Ciudad de M\'exico, Mexico.\\
\small $^{b}$ Escuela Superior de C\'omputo, Instituto Polit\'ecnico Nacional,
Av. Juan de Dios B\'atiz esq. Av. Miguel Oth\'on de Mendiz\'abal, Col. Lindavista,
Delegaci\'on Gustavo A. Madero, C.P. 07738, Ciudad de M\'exico, Mexico.\\

\end{minipage}

\begin{abstract}
In this paper we introduce the bosonic generators of the $sp(4,R)$ algebra and study some of their properties, based on the $SU(1,1)$ and $SU(2)$ group theory. With the developed theory of the $Sp(4,R)$ group, we solve the interaction part of the most general Hamiltonian of a two-level system in two-dimensional geometry in an exact way. As particular cases of this Hamiltonian, we reproduce the solution of earlier problems as the Dirac oscillator and the Jaynes-Cummings model with one and two modes of oscillation.
\end{abstract}

\section{Introduction}

The group methods play an important role in the study and solution of many problems of theoretical physics. For instance, group theory has been applied in high-energy physics, condensed matter, atomic, molecular, and nuclear physics. The rotation group $SO(3)$ was the first group to connect with quantum mechanics through angular momentum formalism. Another important groups in quantum mechanics are the $SU(1,1)$ (which is locally isomorphic to $SO(2,1)$) and $SU(2)$ (which is locally isomorphic to $SO(3)$ and plays a key role in the theory of electron spin). The $SU(1,1)$ and $SU(2)$ groups, together with the so-called potential group $SU_p(1,1)$ can be imbedded into a larger group, $Sp(4,R)$ \cite{Wyb}. The $Sp(4,R)$ group has been extensively studied as can be seen in references \cite{Papa,Castanos85,Castanos86,Castanos87}. One of the most important applications of this group is that it provides a way to calcule transitions from bound states to the continuum for certain potentials, in order to study molecular dissociation. In addition, with this group a unification of the various approaches to one-dimensional potential problems can be achieved \cite{Alhassid}.

In quantum optics, the Jaynes-Cummings model is one of the fundamental theoretical paradigms \cite{Jay,shore}. This model describes the interaction between a two-level atom and a quantized field and is the simplest and completely soluble quantum-mechanical model. The exact solution of this model has been found by using the rotating wave approximation (RWA) \cite{Haroche}. The Jaynes-Cummings model has also been the subject of many generalizations \cite{Koch,Buze,Gou,Nos,Nos1}, besides some of its generalizations are still under study, as shown in references \cite{Lam,Ret,Kop,Sun}.

In addition to the Jaynes-Cummings model, there are other equally important models in quantum optics like the Rabi model \cite{Rabi1,Rabi2}, the Dicke model \cite{Dicke} (also called Tavis-Cummings model \cite{TC}), the $E\bigotimes\epsilon$ Jahn-Teller Hamiltonian \cite{Englman,Reik1}, the modified Jaynes-Cummings model \cite{Kazakov}, among others \cite{Nos2}. Some of these models are particular cases of a general Hamiltonian which is related to the $Sp(4,R)$ group, as can be seen in reference \cite{Koc}. The aim of the present work is to solve exactly the most general Hamiltonian of a two-level system in two-dimensional geometry, by using the $Sp(4,R)$ group theory.

This work is organized as follows. In Section II, we define the generators of the $sp(4,R)$ Lie algebra and the commutation relationships that these operators satisfy. Then, we study the eigenfunctions of the $Sp(4,R)$ group and introduce its coherent states. We obtain the tilting transformation of the generators of this group by using the $SU(1,1)$ and $SU(2)$ group theory. In Section III, we introduce a novelty method based on the theory developed in Section II to solve exactly the interaction part of the most general Hamiltonian of a two-level system in two-dimensional geometry. In Section IV, we study some particular cases of this general Hamiltonian to obtain their energy spectrums. Finally, we give some concluding remarks.

\section{The $Sp(4,R)$ group and its similarity transformations}

In this Section, we give a brief introduction to the main realizations of the $su(2)$ and $su(1,1)$ Lie algebras based on the bosonic annihilation $\hat{a}$, $\hat{b}$ and creation $\hat{a}^{\dag}$, $\hat{b}^{\dag}$ operators. We shall obtain the commutation relations of the $sp(4,R)$ algebra which satisfy the operators of the $su(2)$ and $su(1,1)$ Lie algebras realizations. Also, we compute the transformations of the operators which belong to $sp(4,R)$ algebra by using the similarity transformations of the $SU(2)$ and $SU(1,1)$ displacement operator.

\subsection{The $su(2)$ and $su(1,1)$ Lie algebras}

The study of the $Sp(4,R)$ group starts with the introduction of the $su(1,1)$ and $su(2)$ Lie algebras, which satisfy the following commutation relations
\begin{eqnarray}
[K_{0},K_{\pm}]=\pm K_{\pm},\quad\quad [K_{-},K_{+}]=2K_{0},\label{algebra1}
\end{eqnarray}
\begin{eqnarray}
[J_{0},J_{\pm}]=\pm J_{\pm},\quad\quad [J_{+},J_{-}]=2J_{0}.\label{algebra2}
\end{eqnarray}
In these expressions, the operators $K_{\pm}$, $K_0$ are the generators of the $su(1,1)$ Lie algebra and the operators $J_\pm$ and $J_{0}$ are the generators of the $su(2)$ Lie algebra. The Casimir operators $K^2$ and $J^{2}$ for these algebras have the form
\begin{equation}
K^2=K_0^2-\frac{1}{2}\left(K_+K_-+K_-K_+\right), \quad\quad J^{2}=J_0^2+\frac{1}{2}\left(J_+J_-+J_-J_+\right).
\end{equation}
which satisfy $[K^{2},K_{\pm}]=[K^{2},K_{0}]=0$ and $[J^{2},J_{\pm}]=[J^{2},J_{0}]=0$,

Now, the discrete representation of the $su(1,1)$ Lie algebra is given by
\begin{equation}
K_{+}|k,n\rangle=\sqrt{(n+1)(2k+n)}|k,n+1\rangle,\label{k+n}
\end{equation}
\begin{equation}
K_{-}|k,n\rangle=\sqrt{n(2k+n-1)}|k,n-1\rangle,\label{k-n}
\end{equation}
\begin{equation}
K_{0}|k,n\rangle=(k+n)|k,n\rangle,\label{k0n}
\end{equation}
\begin{equation}
K^2|k,n\rangle=k(k-1)|k,n\rangle,\label{Cas}
\end{equation}
where $|k,n\rangle$ are the eigenstates of $K_{0}$ and $K^{2}$, being $k$ the Bargmann index and $n$ a non-negative integer. The state $|k,n\rangle$ can be generated from the ``ground" state or the lowest state $|k,0\rangle$ according to
\begin{equation}
|k,n\rangle=\left[\frac{\Gamma(2k)}{n!\Gamma(2k+n)}\right]^{1/2}(K_{+})^{n}|k,0\rangle.
\end{equation}
The set of states $|k,n\rangle$,
\begin{equation}
S_{su(1,1)}=\{|k,n\rangle \quad|\quad n=0,1,2,...;\quad k= const.\}
\end{equation}
becomes a complete orthonormal basis
\begin{equation}
\sum_{n=0}^{\infty}|k,n\rangle\langle k,n|=1, \quad\quad \langle k,n|k,n\rangle=\delta_{k,n}.
\end{equation}

On the other hand, the discrete representation of the $su(2)$ Lie algebra is given by
\begin{equation}
J_{+}|j,\mu \rangle=\sqrt{(j-\mu)(j+\mu+1)}|j,\mu+1 \rangle,\label{j+m}
\end{equation}
\begin{equation}
J_{-}|j,\mu \rangle=\sqrt{(j+\mu)(j-\mu+1)}|j,\mu-1 \rangle,\label{j-m}
\end{equation}
\begin{equation}
J_{0}|j,\mu \rangle=\mu|j,\mu \rangle,\label{j0m}
\end{equation}
\begin{equation}
J^2|j,\mu \rangle=j(j+1)|j,\mu \rangle,
\end{equation}
where $J_{-}|j,-j \rangle=J_{+}|j,j\rangle=0$. Also, any state can be obtained in terms of the ``ground" state by the relationship
\begin{equation}
|j,\mu\rangle=\left[\frac{(j-\mu)!}{(2j)!(j+\mu)!}\right]^{1/2}(J_{+})^{j+\mu}|j,-j\rangle.
\end{equation}
In this case $j=0,1/2,1,3/2,2,...$ and $\mu=-j,-j+1,...,j-1,j$. The set
\begin{equation}
S_{su(2)}=\{|j,\mu\rangle\quad|\quad \mu=-j,-j+1,...,j-1,j;\quad j= const.\}
\end{equation}
becomes a complete orthonormal basis that satisfies
\begin{equation}
\sum_{\mu=-j}^{j}|j,\mu\rangle\langle j,\mu|=1, \quad\quad \langle j,\mu|j,\mu\rangle=\delta_{j,\mu}.
\end{equation}

As it is well known, the bosonic annihilation $\hat{a}$, $\hat{b}$ and creation $\hat{a}^{\dag}$, $\hat{b}^{\dag}$ operators obey the commutation relations
\begin{equation}
[\hat{a},\hat{a}^{\dag}]=[\hat{b},\hat{b}^{\dag}]=1,
\end{equation}
\begin{equation}
[\hat{a},\hat{b}]=[\hat{a}^{\dag},\hat{b}^{\dag}]=[\hat{a}^{\dag},\hat{b}]=[\hat{a},\hat{b}^{\dag}]=0.\label{boson}
\end{equation}
These operators can be used to construct some of the $su(2)$ and $su(1,1)$ algebra realizations. Thus, with the bilinear products $\hat{a}^{\dag}\hat{a}$, $\hat{b}^{\dag}\hat{b}$, $\hat{a}^{\dag}\hat{b}$ and $\hat{b}^{\dag}\hat{a}$ we can construct the $su(2)$ Lie algebra realization by introducing the four operators
\begin{equation}
J_+=\hat{a}^{\dag}\hat{b},\quad\quad J_-=\hat{b}^{\dag}\hat{a},\quad\quad J_0=\frac{1}{2}(\hat{a}^{\dag}\hat{a}-\hat{b}^{\dag}\hat{b}),\quad\quad
J^{2}=\frac{1}{4}N(N+2),\label{su2}
\end{equation}
where $N=\hat{a}^{\dag}\hat{a}+\hat{b}^{\dag}\hat{b}$.

In the same sense, with the operators $\hat{a}^{\dag}\hat{a}$, $\hat{b}^{\dag}\hat{b}$, $\hat{a}^{\dag}\hat{b}^{\dag}$, $\hat{b}\hat{a}$, $\hat{a}^{\dag}{}^{2}$ and $\hat{a}^{2}$ we can construct two different realizations of the $su(1,1)$ Lie algebra with the set of operators
\begin{equation}
K_{+}^{(ab)}=\hat{a}^{\dag}\hat{b}^{\dag},\quad\quad K_{-}^{(ab)}=\hat{b}\hat{a},\quad\quad K_{0}^{(ab)}=\frac{1}{2}(\hat{a}^{\dag}\hat{a}+\hat{b}^{\dag}\hat{b}+1),\quad
\quad K_{(ab)}^{2}=J_{0}^{2}-\frac{1}{4}, \label{su11ab}
\end{equation}
and
\begin{equation}
K_{+}^{(a)}=\frac{1}{2}\hat{a}^{\dag}{}^{2},\quad\quad K_{-}^{(a)}=\frac{1}{2}\hat{a}^{2},\quad\quad K_{0}^{(a)}=\frac{1}{2}\left(\hat{a}^{\dag}\hat{a}+\frac{1}{2}\right),\quad\quad K_{(a)}^{2}=-\frac{3}{16}. \label{su11a}
\end{equation}

Notice that for the second realization of the $su(1,1)$ Lie algebra, the Casimir operator $K_{(a)}^{2}$ is constant. This implies that the
Bargmann index $k$ only can take the values $k=\frac{1}{4}$ and $k=\frac{3}{4}$. Now, if we define the set of operators
\begin{equation}
K_{+}^{(b)}=\frac{1}{2}\hat{b}^{\dag}{}^{2},\quad\quad K_{-}^{(b)}=\frac{1}{2}\hat{b}^{2},\quad\quad K_{0}^{(b)}=\frac{1}{2}\left(\hat{b}^{\dag}\hat{b}+\frac{1}{2}\right)\quad\quad K_{(b)}^{2}=-\frac{3}{16},\label{su11b}
\end{equation}
we can construct another interesting realization of the $su(1,1)$ Lie algebra, generated by the operators
\begin{equation}
K_{+}=\frac{1}{2}\left(\hat{a}^{\dag}{}^{2}+\hat{b}^{\dag}{}^{2}\right),\quad\quad K_{-}=\frac{1}{2}\left(\hat{a}^{2}+\hat{b}^{2}\right),\quad\quad K_{0}^{(ab)}=\frac{1}{2}\left(\hat{a}^{\dag}\hat{a}+\hat{b}^{\dag}\hat{b}+1\right),
\end{equation}
\begin{equation}
K^{2}=\left(K_{0}^{(ab)}\right)^{2}-\frac{1}{2}\left(\{K_{+}^{(a)},K_{-}^{(a)}\}+\{K_{+}^{(b)},K_{-}^{(b)}\}\right)-\frac{1}{4}\left(J_{+}^{2}+J_{-}^{2}\right).
\end{equation}

\subsection{The generators of the $Sp(4,R)$ group}

With the above background, we can introduce the most general quadratic form in terms of the boson operators $\hat{a}$ and $\hat{b}$, which is formed by the following $10$ operators \cite{Castanos85,Alhassid}
\begin{equation}
K_{+}^{(ab)}, \hspace{0.2cm} K_{+}^{(a)}, \hspace{0.2cm}K_{+}^{(b)}, \hspace{0.2cm}J_{+}, \hspace{0.2cm}K_{-}^{(ab)}, \hspace{0.2cm}K_{-}^{(a)},
\hspace{0.2cm}K_{-}^{(b)}, \hspace{0.2cm}J_{-}, \hspace{0.2cm}K_{0}^{(ab)}, \hspace{0.2cm}J_{0}. \label{sp4r}
\end{equation}
These operators close the symplectic algebra $sp(4,R)$. The $Sp(4,R)$ group contains as subgroups the bound state group $SU(2)$ and its invariant $K_{0}^{(ab)}$, and the scattering state group $SU(1,1)$ and its invariant $J_{0}$. Thus, this group provides a unified framework of both bound and scattering states and therefore, a unified treatment of the various approaches to the solution of problems which are described by the $su(2)$ and $su(1,1)$ algebras.

The $10$ operators of equation (\ref{sp4r}) belonging to the $Sp(4,R)$ group obey the following commutation relations
\begin{equation}
[K_{0}^{(ab)},K_{\pm}^{(i)}]=\pm K_{\pm}^{(i)},\quad\quad [K_{-}^{(i)},K_{-}^{(i)}]=[K_{+}^{(i)},K_{+}^{(i)}]=0,\quad\quad i=a,b,ab,
\end{equation}
\begin{equation}
[K_{0}^{(ab)},J_{0}]=[K_{0}^{(ab)},J_{\pm}]=0.
\end{equation}
The rest of the commutation relations between these generators are given in Table 1.

\begin{table}[!htbp]
\begin{center}
\begin{tabular}{|c|c|c|c|c|c|c|}
\hline
&$K_{-}^{(a)}$&$K_{-}^{(b)}$&$K_{-}^{(ab)}$&$J_{+}$&$J_{0}$&$J_{-}$\\\hline \hline
$K_{+}^{(a)}$&$-\frac{1}{2}\left(K_{0}^{(ab)}+J_{0}\right)$&$0$&$-J_{+}$&$0$&$-K_{+}^{(a)}$&$-K_{+}^{(ab)}$\\
$K_{+}^{(b)}$&$0$&$-\frac{1}{2}\left(K_{0}^{(ab)}-J_{0}\right)$&$-J_{-}$&$-K_{+}^{(ab)}$&$-K_{+}^{(b)}$&$0$\\
$K_{+}^{(ab)}$&$-J_{-}$&$-J_{+}$&$-2K_{0}^{(ab)}$&$-2K_{+}^{(a)}$&$0$&$-2K_{+}^{(b)}$\\
$J_{+}$&$-K_{-}^{(ab)}$&$0$&$-2K_{-}^{(b)}$&$0$&$-J_{+}$&$2J_{0}$\\
$J_{0}$&$-K_{-}^{(a)}$&$K_{-}^{(b)}$&$0$&$J_{+}$&$0$&$-J_{-}$\\
$J{-}$&$0$&$-K_{-}^{ab}$&$-2K_{-}^{(a)}$&$-2J_{0}$&$J_{-}$&$0$\\\hline
\end{tabular}
\caption{Shows in a condensed way some of the commutation relations of the $Sp(4,R)$ group generators.}
\end{center}
\end{table}
Here, each term that appears in the body of the Table 1 is the commutator between one of the terms that appear in the first column on the left-hand ($K_{+}^{(a)}$, $K_{+}^{(b)}$, $K_{+}^{(ab)}$, $J_{+}$, $J_{0}$, $J_{-}$) and one of the terms in the first row of the top ($K_{-}^{(a)}$, $K_{-}^{(b)}$, $K_{-}^{(ab)}$, $J_{+}$, $J_{0}$, $J_{-}$). For example, $[K_{+}^{(ab)},K_{-}^{(a)}]=-J_{-}$.

Moreover, the generators of the $Sp(4,R)$ group can be written in a representation defined in terms of $4\times4$ matrices as
\begin{equation}
J_{0}=\begin{pmatrix}\sigma_{0}&0\\0&\sigma_{0}\\ \end{pmatrix},\quad\quad J_{+}=\begin{pmatrix}\sigma_{+}&0\\0&-\sigma_{+}\\ \end{pmatrix},\quad\quad J_{-}=\begin{pmatrix}\sigma_{-}&0\\0&-\sigma_{-}\\ \end{pmatrix},
\end{equation}
\begin{equation}
 K_{0}^{(ab)}=\frac{1}{2}\begin{pmatrix}I&0\\0&-I\\\end{pmatrix},\quad\quad K_{+}^{(ab)}=\begin{pmatrix}
 0&0\\i\sigma_{0}^{2}&0\\\end{pmatrix},\quad\quad K_{-}^{(ab)}=\begin{pmatrix}0&-i\sigma_{0}^{2}\\0&0\\\end{pmatrix},
\end{equation}
\begin{equation}
K_{+}^{(a)}=\begin{pmatrix}0&0\\i\sigma_{+}\sigma_{0}&0\\\end{pmatrix},\quad\quad K_{-}^{(a)}=\begin{pmatrix}0&-i\sigma_{0}\sigma_{-}\\0&0\\\end{pmatrix},
\end{equation}
\begin{equation}
K_{+}^{(b)}=\begin{pmatrix}0&0\\i\sigma_{0}\sigma_{-}&0\\\end{pmatrix},\quad\quad K_{-}^{(b)}=\begin{pmatrix}0&-i\sigma_{+}\sigma_{0}\\0&0\\\end{pmatrix},
\end{equation}
where all submatrices are $2\times2$, $I$ is the identity matrix and $\sigma_{+},\sigma_{-},\sigma_{0}$ are the Pauli spin matrices
\begin{equation}
\sigma_{+}=\begin{pmatrix}0&1\\0&0\\\end{pmatrix},\quad\quad\sigma_{-}=\begin{pmatrix}0&0\\1&0\\\end{pmatrix},\quad\quad\sigma_{0}\begin{pmatrix}1&0\\0&-1\\\end{pmatrix}.
\end{equation}
It is easy to show that in this matrix representation the generators satisfy all commutation relations of Table 1.

These 10 generators of $Sp(4,R)$ can be divided into three subsets: lowering, raising and  weight operators. In this realization, the set of raising operators is given by the operators $\{K_{+}^{(ab)}, K_{+}^{(a)}, K_{+}^{(b)}, J_{+}\}$, the set of lowering operators is given by $\{K_{-}^{(ab)}, K_{-}^{(a)}, K_{-}^{(b)}, J_{-}\},$ and the set of weight operators is given by  $\{K_{0}^{(ab)}, J_{0}\}$.

In general, the lowest state of the $Sp(4,R)$ group is characterized by a state $|\omega\rangle$ which satisfies the equations \cite{Castanos86}
\begin{equation}
K_{-}^{(i)}|\omega\rangle=0,\quad\quad J_{-}|\omega\rangle=0, \quad\quad\quad i=a,b,ab,
\end{equation}
\begin{equation}
K_{0}^{(ab)}|\omega\rangle=k|\omega\rangle,\quad\quad J_{0}|\omega\rangle=-j|\omega\rangle.
\end{equation}

Now, since we are using the realization of the $Sp(4,R)$ group given in terms of the creation $\hat{a}^{\dag}$, $\hat{b}^{\dag}$ and annihilation $\hat{a}$, $\hat{b}$ operators, we can use the usual number states of the two-mode field $|n,m\rangle$. These states satisfy the following relationships with respect to the raising and lowering operators
\begin{equation}
K_{+}^{(ab)}|n,m\rangle\longrightarrow |n+2,m\rangle,\quad\quad K_{+}^{(a)}|n,m\rangle\longrightarrow|n+2,m+2\rangle,\quad\quad K_{+}^{(b)}|n,m\rangle\longrightarrow|n+2,m-2\rangle,\label{+}
\end{equation}
\begin{equation}
K_{-}^{(ab)}|n,m\rangle\longrightarrow|n-2,m\rangle,\quad\quad K_{-}^{(a)}|n,m\rangle\longrightarrow|n-2,m-2\rangle,\quad\quad K_{-}^{(b)}|n,m\rangle\longrightarrow|n-2,m+2\rangle,\label{-}
\end{equation}
\begin{equation}
J_{+}|n,m\rangle\longrightarrow|n,m+2\rangle\quad\quad J_{-}|n,m\rangle\longrightarrow|n,m-2\rangle.\label{J+-}
\end{equation}
The states $|n,m\rangle$ can be represented by states of the $SU(1,1)$ and $SU(2)$ groups as $|k,m_{0}\rangle$ and $|j,\mu\rangle$, with the following relationship between group numbers and quantum numbers
\begin{equation}
k=\frac{m+1}{2},\quad\quad m_{0}=\frac{n-m}{2},\quad\quad j=\frac{n}{2},\quad\quad \mu=\frac{m}{2}.
\end{equation}
The lowest states of the $SU(1,1)$ and $SU(2)$ groups are given respectively by $|k,0\rangle$ and $|j,-j\rangle$. Therefore, for the state $|k,0\rangle$ we have that  $k=(m+1)/2$ and $n=m$, and for the state $|j,-j\rangle$  we have that $n=-m$. Thus, the lowest state of the $Sp(4,R)$ group is given by
\begin{equation}
|\omega\rangle=|k,j,-j\rangle =|k,0\rangle\bigotimes|j,-j\rangle.\label{lowsp4}
\end{equation}

A complete set of states is given by applying powers of the raising generators to the lowest state
$|k,j,-j\rangle$. Therefore, as can be seen in Ref. \cite{Castanos87} a general state of the $Sp(4,R)$ group can be obtained as
\begin{equation}
|N,M,\mu,\sigma\rangle=\left( K^{(a)}_{+}\right)^{(N+M-\mu-\sigma)/2}\left(K^{(ab)}_{+}\right)^{\mu}\left(K^{(b)}_{+}\right)^{(N-M-\mu+\sigma)/2}\left(J_{+}\right)^{j+\sigma}|K,j,-j\rangle,\label{ketsp4r}
\end{equation}
where  $N, M, \mu, \sigma$ are the $Sp(4,R)$ group numbers. Also, it can be shown that (\ref{ketsp4r}) is an eigenstate of the operators $N$ and $J_{0}$
\begin{equation}
N|N,M,\mu,\sigma\rangle=(N+k)|N,M,\mu,\sigma\rangle\quad\quad J_{0}|N,M,\mu,\sigma\rangle=M|N,M,\mu,\sigma\rangle.
\end{equation}
These expressions are the relations (\ref{k0n}) and (\ref{j0m}) of the $SU(1,1)$ and $SU(2)$ group states, respectively.

The coherent states of $Sp(4,R)$ group can be introduced as the action of the displacement operators for each realization of the $SU(1,1)$ and $SU(2)$ groups on the lowest state \cite{Castanos87}
\begin{align}
|\xi_{a},\xi_{b},\xi,\chi,\omega\rangle=&D_{su(1,1)}(\xi_{a})D_{su(1,1)}(\xi_{b})D_{su(1,1)}(\xi)D_{su(2)}(\chi)|K,j,-j\rangle\\ \nonumber =&D_{su(1,1)}(\xi_{a})D_{su(1,1)}(\xi_{b})|\zeta_{1}\rangle|\zeta_{2}\rangle.
\end{align}
In this expression, the states $|\zeta_{1}\rangle$ and $|\zeta_{2}\rangle$ result to be the $SU(1,1)$ and $SU(2)$ Perelomov coherent states of the two-dimensional harmonic oscillator, for the realizations given in equations (\ref{su2}) and (\ref{su11ab}) (see the Appendices).

\subsection{The similarity transformations of the $Sp(4,R)$ group generators}

Considering the realizations of equations (\ref{su11a}) and (\ref{su11b}) of the $su(1,1)$ Lie algebra, we can introduce
the following exponential operator
\begin{align}
D(\xi)_{a,b}&= \exp[\xi_{a}K_{+}^{(a)}-\xi_{a}^{*}K_{-}^{(a)}+\xi_{b}K_{+}^{(b)}-\xi_{b}^{*}K_{-}^{(b)}]\nonumber\\&=\exp[\xi_{a}K_{+}^{(a)}-\xi_{a}^{*}K_{-}^{(a)}]\exp[\xi_{b}K_{+}^{(b)}-\xi_{b}^{*}K_{-}^{(b)}]\nonumber\\&=D(\xi_{a})D(\xi_{b}),\label{su11xsu11}
\end{align}
where the operators $D(\xi_{a})$ and $D(\xi_{b})$ are the $SU(1,1)$ displacement operators with different modes of oscillation (see Appendix A).
With this operator we can compute the similarity transformation of the $Sp(4,R)$ generators as follows
\begin{align}
D_{a,b}^{\dag}J_{0}D_{a,b}&=(\cosh(2|\xi_{a}|)+\cosh(2|\xi_{b}|))J_{0}+(\cosh(2|\xi_{a}|)-\cosh(2|\xi_{b}|))K_{0}^{(ab)}
-\frac{\sinh(2|\xi_{a}|)}{|\xi_{a}|}\left(\xi_{a}K_{+}^{(a)}+\xi_{a}^{*}K_{-}^{(a)}\right)\nonumber\\&
+\frac{\sinh(2|\xi_{b}|)}{|\xi_{b}|}\left(\xi_{b}K_{+}^{(b)}+\xi_{b}^{*}K_{-}^{(b)}\right),
\end{align}
\begin{align}
&D_{a,b}^{\dag}J_{+}D_{a,b}=J_{+}\cosh_{a}\cosh_{b}-K_{-}^{(ab)}\frac{\xi_{a}^{*}}{|\xi_{a}|}\cosh_{b}\sinh_{a}-K_{+}^{(ab)}\frac{\xi_{b}}{|\xi_{b}|}\cosh_{a}\sinh _{b}+J_{-}\frac{\xi_{a}^{*}\xi_{b}}{|\xi_{a}||\xi_{b}|}\sinh_{a}\sinh_{b},
\end{align}
\begin{align}
D_{a,b}^{\dag}J_{-}D_{a,b}=J_{-}\cosh_{a}\cosh_{b}-K_{+}^{(ab)}\frac{\xi_{a}}{|\xi_{a}|}\cosh_{b}\sinh_{a}-K_{-}^{(ab)}\frac{\xi_{b}^{*}}{|\xi_{b}|}\cosh_{a}\sinh _{b}+J_{+}\frac{\xi_{a}\xi_{b}^{*}}{|\xi_{a}||\xi_{b}|}\sinh_{a}\sinh_{b},
\end{align}
\begin{align}
D_{a,b}^{\dag}K_{0}^{(ab)}D_{a,b}&=(\cosh(2|\xi_{a}|)-\cosh(2|\xi_{b}|))J_{0}+(\cosh(2|\xi_{a}|)+\cosh(2|\xi_{b}|))K_{0}^{(ab)}
-\frac{\sinh(2|\xi_{a}|)}{|\xi_{a}|}\left(\xi_{a}K_{+}^{(a)}+\xi_{a}^{*}K_{-}^{(a)}\right)\nonumber\\&
-\frac{\sinh(2|\xi_{b}|)}{|\xi_{b}|}\left(\xi_{b}K_{+}^{(b)}+\xi_{b}^{*}K_{-}^{(b)}\right)-1,
\end{align}
\begin{align}
D_{a,b}^{\dag}K_{-}^{(ab)}D_{a,b}=K_{-}^{(ab)}\cosh_{a}\cosh_{b}-J_{+}\frac{\xi_{a}}{|\xi_{a}|}\cosh_{b}\sinh_{a}-J_{-}\frac{\xi_{b}}{|\xi_{b}|}\cosh_{a}\sinh _{b}+K_{+}^{(ab)}\frac{\xi_{a}\xi_{b}}{|\xi_{a}||\xi_{b}|}\sinh_{a}\sinh_{b},
\end{align}
\begin{align}
D_{a,b}^{\dag}K_{+}^{(ab)}D_{a,b}=K_{+}^{(ab)}\cosh_{a}\cosh_{b}-J_{-}\frac{\xi_{a}^{*}}{|\xi_{a}|}\cosh_{b}\sinh_{a}-J_{+}\frac{\xi_{b}^{*}}{|\xi_{b}|}\cosh_{a}\sinh _{b}+K_{-}^{(ab)}\frac{\xi_{a}^{*}\xi_{b}^{*}}{|\xi_{a}||\xi_{b}|}\sinh_{a}\sinh_{b},
\end{align}
where $\cosh_{a}=\cosh(|\xi_{a}|)$, $\sinh_{a}=\sinh(|\xi_{a}|)$, $\cosh_{b}=\cosh(|\xi_{b}|)$, $\sinh_{b}=\sinh(|\xi_{b}|)$ and $\xi_{a}$, $\xi_{b}$ are complex constants.

The other important similarity transformations of the $Sp(4,R)$ generators can be obtained in terms of the $SU(1,1)$ displacement operator represented in the Jordan-Schwinger realization of equation (\ref{su11ab})
\begin{equation}
D(\xi)= \exp[\xi K_{+}^{(ab)}-\xi^{*}K_{-}^{(ab)}].\label{Dxi}
\end{equation}
Thus, with this operator we obtain the following results
\begin{align}
&D^{\dag}K_{-}^{(a)}D=\frac{K_{-}^{(a)}}{2}\left(\cosh(2|\xi|)+1\right)+K_{+}^{(b)}\frac{\xi}{2\xi^{*}}\left(\cosh(2|\xi|)-1\right)-J_{-}\frac{\xi}{2|\xi|}\sinh(2|\xi|),\label{trsu11ab1}\\&
D^{\dag}K_{+}^{(a)}D=\frac{K_{+}^{(a)}}{2}\left(\cosh(2|\xi|)+1\right)+K_{-}^{(b)}\frac{\xi^{*}}{2\xi}\left(\cosh(2|\xi|)-1\right)-J_{+}\frac{\xi^{*}}{2|\xi|}\sinh(2|\xi|),\\&
D^{\dag}K_{-}^{(b)}D=\frac{K_{-}^{(b)}}{2}\left(\cosh(2|\xi|)+1\right)+K_{+}^{(a)}\frac{\xi}{2\xi^{*}}\left(\cosh(2|\xi|)-1\right)-J_{+}\frac{\xi}{2|\xi|}\sinh(2|\xi|),\\&
D^{\dag}K_{+}^{(b)}D=\frac{K_{+}^{(b)}}{2}\left(\cosh(2|\xi|)+1\right)+K_{-}^{(a)}\frac{\xi^{*}}{2\xi}\left(\cosh(2|\xi|)-1\right)-J_{-}\frac{\xi^{*}}{2|\xi|}\sinh(2|\xi|),\\&
D^{\dag}J_{+}D=J_{+}\cosh(2|\xi|)-K_{+}^{(a)}\frac{\xi}{|\xi|}\sinh(2|\xi|)-K_{-}^{(b)}\frac{\xi^{*}}{|\xi|}\sinh(2|\xi|),\\&
D^{\dag}J_{-}D=J_{-}\cosh(2|\xi|)-K_{-}^{(a)}\frac{\xi^{*}}{|\xi|}\sinh(2|\xi|)-K_{+}^{(b)}\frac{\xi}{|\xi|}\sinh(2|\xi|),\\&
D^{\dag}J_{0}D=J_{0},\label{trsu11ab}
\end{align}
where $\xi$ is a complex parameter.

The last similarity transformations of the $Sp(4,R)$ generators is obtained by using the $SU(2)$ displacement operator based on the realization of equation (\ref{su2}):
\begin{equation}
D(\chi)= \exp[\chi J_{+}-\chi^{*}J_{-}].
\end{equation}
We obtain
\begin{align}
&D^{\dag}K_{-}^{(a)}D=\frac{K_{-}^{(a)}}{2}\left(\cos(2|\chi|)+1\right)-K_{-}^{(b)}\frac{\chi}{2\chi^{*}}\left(\cos(2|\chi|)-1\right)-K_{-}^{(ab)}\frac{\chi}{2|\chi|}\sin(2|\xi|),\\&
D^{\dag}K_{+}^{(a)}D=\frac{K_{+}^{(a)}}{2}\left(\cos(2|\chi|)+1\right)-K_{+}^{(b)}\frac{\chi^{*}}{2\chi}\left(\cos(2|\chi|)-1\right)-K_{+}^{(ab)}\frac{\chi^{*}}{2|\chi|}\sin(2|\xi|),\\&
D^{\dag}K_{-}^{(b)}D=\frac{K_{-}^{(b)}}{2}\left(\cos(2|\chi|)+1\right)-K_{-}^{(a)}\frac{\chi^{*}}{2\chi}\left(\cos(2|\chi|)-1\right)+K_{-}^{(ab)}\frac{\chi^{*}}{2|\chi|}\sin(2|\chi|),\\&
D^{\dag}K_{+}^{(b)}D=\frac{K_{+}^{(b)}}{2}\left(\cos(2|\chi|)+1\right)-K_{+}^{(a)}\frac{\chi}{2\chi^{*}}\left(\cos(2|\chi|)-1\right)+K_{+}^{(ab)}\frac{\chi}{2|\chi|}\sin(2|\chi|),\\&
D^{\dag}K_{+}^{(ab)}D=K_{+}^{(a)}\frac{\chi}{|\chi|}\sin(2|\chi|)-K_{+}^{(b)}\frac{\chi^{*}}{|\chi|}\sin(2|\chi|)+K_{+}^{(ab)}\cos(2|\chi|),\\&
D^{\dag}K_{-}^{(ab)}D=K_{-}^{(a)}\frac{\chi^{*}}{|\chi|}\sin(2|\chi|)-K_{-}^{(b)}\frac{\chi}{|\chi|}\sin(2|\chi|)+K_{-}^{(ab)}\cos(2|\chi|),\\&
D^{\dag}K_{0}^{(ab)}D=K_{0}^{(ab)},
\end{align}
where again $\chi$ is a complex parameter.

We observe from these transformations of the $Sp(4,R)$ group generators that there is a connection between the $SU(1,1)$ and $SU(2)$ groups. This connection is made through the main realizations of the $su(2)$ and $su(1,1)$ Lie algebras, which are based on the bosonic annihilation $\hat{a}$, $\hat{b}$ and creation $\hat{a}^{\dag}$, $\hat{b}^{\dag}$ operators. In the next Section, we shall use this connection to solve exactly a general Hamiltonian related to the $su(1,1)$ and $su(2)$ algebras, as well as the $sp(4,R)$ algebra.

\section{The general Hamiltonian of a two-level system in two-dimensional geometry}

The most general form of the Hamiltonian that describe a two-level system for a single particle with spin-$1/2$, in two dimensional geometry is given by \cite{Koc}
\begin{equation}
\hat{H}=\hbar\omega_{1}\hat{a}^{\dag}\hat{a}+\hbar\omega_{2}\hat{b}^{\dag}\hat{b}+\frac{\hbar\omega_{0}}{2}\sigma_{0}+(\kappa_{1}\hat{a}+\kappa_{2}\hat{a}^{\dag}+\kappa_{3}\hat{b}+\kappa_{4}\hat{b}^{\dag})\sigma_{+}+(\gamma_{1}\hat{a}+\gamma_{2}\hat{a}^{\dag}+\gamma_{3}\hat{b}+\gamma_{4}\hat{b}^{\dag})\sigma_{-},\label{H}
\end{equation}
where $\omega_{i}$, $\kappa_{i}$ and $\gamma_{i}$ are physical constants, $\sigma_{0}$ and $\sigma_{\pm}$ are usual Pauli matrices, $\hat{a}$, $\hat{b}$ and  $\hat{a}^{\dag}$, $\hat{b}^{\dag}$ are the bosonic annihilation and creation operators. This Hamiltonian can be written as $\hat{H}=\hat{H}_{0}+\hat{H}_{I}$, with
\begin{equation}
\hat{H}_{0}=\hbar\omega_{1}\left(\hat{a}^{\dag}\hat{a}+\frac{\sigma_{0}}{2}\right)+\hbar\omega_{2}\left(\hat{b}^{\dag}\hat{b}+\frac{\sigma_{0}}{2}\right),
\end{equation}
\begin{equation}
\hat{H}_{I}=\hbar\left(\frac{\omega_{0}-\omega_{1}-\omega_{2}}{2}\right)\sigma_{0}+(\kappa_{1}\hat{a}+\kappa_{2}\hat{a}^{\dag}+\kappa_{3}\hat{b}+\kappa_{4}\hat{b}^{\dag})\sigma_{+}+(\gamma_{1}\hat{a}+\gamma_{2}\hat{a}^{\dag}+\gamma_{3}\hat{b}+\gamma_{4}\hat{b}^{\dag})\sigma_{-}.\label{HI}
\end{equation}

We are interested in solving the eigenvalues equation of the interaction Hamiltonian
\begin{equation}
\hat{H}_{I}|\varphi\rangle=E|\varphi\rangle,
\end{equation}
where $|\varphi\rangle$ is a two component spinor and $E$ is its eigenvalue. The coupled equations for the spinor components $|\varphi_{1}\rangle$ and $|\varphi_{2}\rangle$ are
\begin{equation}
\left(\kappa_{1}\hat{a}+\kappa_{2}\hat{a}^{\dag}+\kappa_{3}\hat{b}+\kappa_{4}\hat{b}^{\dag}\right)|\varphi_2\rangle=\left(E-\hbar\Delta\omega\right)|\varphi_1\rangle,\label{1}
\end{equation}
\begin{equation}
\left(\gamma_{1}\hat{a}+\gamma_{2}\hat{a}^{\dag}+\gamma_{3}\hat{b}+\gamma_{4}\hat{b}^{\dag}\right)|\varphi_1\rangle=\left(E+\hbar\Delta\omega\right)|\varphi_2\rangle,\label{2}
\end{equation}
where $\Delta\omega=\frac{\omega_{0}-\omega_{1}-\omega_{2}}{2}$.

Uncoupling these relations we find that the equations for $|\varphi_1\rangle$ and $|\varphi2\rangle$ result to be
\begin{equation}
\left(\kappa_{1}\hat{a}+\kappa_{2}\hat{a}^{\dag}+\kappa_{3}\hat{b}+\kappa_{4}\hat{b}^{\dag}\right)\left(\gamma_{1}\hat{a}+\gamma_{2}\hat{a}^{\dag}+\gamma_{3}\hat{b}+\gamma_{4}\hat{b}^{\dag}\right)|\varphi_1\rangle=\left(E^{2}-(\hbar\Delta\omega)^{2}\right)|\varphi_1\rangle,\label{un1}
\end{equation}
\begin{equation}
\left(\gamma_{1}\hat{a}+\gamma_{2}\hat{a}^{\dag}+\gamma_{3}\hat{b}+\gamma_{4}\hat{b}^{\dag}\right)\left(\kappa_{1}\hat{a}+\kappa_{2}\hat{a}^{\dag}+\kappa_{3}\hat{b}+\kappa_{4}\hat{b}^{\dag}\right)|\varphi_2\rangle=\left(E^{2}-(\hbar\Delta\omega)^{2}\right)|\varphi_2\rangle.\label{un2}
\end{equation}
Since these two equations have the same mathematical structure, we will focus only on the equation for $|\varphi_1\rangle$. Then, multiplying each term on the left-hand of the uncoupled equation for $|\varphi_1\rangle$ and using the $Sp(4,R)$ group generators of equation (\ref{sp4r}), we can write the equation (\ref{un1}) as
\begin{align}
&\left[2\alpha_{1}K_{-}^{(a)}+2\alpha_{2}K_{+}^{(a)}+2\alpha_{3}K_{-}^{(b)}+2\alpha_{4}K_{+}^{(b)}+(\alpha_{5}+\alpha_{6})K_{0}^{(ab)}+\alpha_{7}K_{-}^{(ab)}+\alpha_{10}K_{+}^{(ab)}\right.\nonumber\\&\left.+(\alpha_{5}-\alpha_{6})J_{0}+\alpha_{8}J_{-}+\alpha_{9}J_{+}+\alpha_{11}-\frac{\alpha_{5}+\alpha_{6}}{2}\right]|\varphi_1\rangle=(E^{2}-(\hbar\Delta\omega)^{2})|\varphi_1\rangle,\label{Unc1}
\end{align}
where we have defined the $\alpha$-constants as
\begin{align}
&\alpha_{1}=\kappa_{1}\gamma_{1},\quad\quad\alpha_{2}=\kappa_{2}\gamma_{2},\quad\quad\alpha_{3}=\kappa_{3}\gamma_{3},\quad\quad\alpha_{4}=\kappa_{4}\gamma_{4},\quad\quad
\alpha_{5}=\kappa_{1}\gamma_{2}+\kappa_{2}\gamma_{1},\nonumber\\&\alpha_{6}=\kappa_{4}\gamma_{3}+\kappa_{3}\gamma_{4},\quad\quad\alpha_{7}=\kappa_{1}\gamma_{3}+\kappa_{3}\gamma_{1},\quad\quad\alpha_{8}=\kappa_{1}\gamma_{4}+\kappa_{4}\gamma_{1},\nonumber\\&
\alpha_{9}=\kappa_{2}\gamma_{3}+\kappa_{3}\gamma_{2},\quad\quad\alpha_{10}=\kappa_{2}\gamma_{4}+\kappa_{4}\gamma_{2},\quad\quad\alpha_{11}=\kappa_{1}\gamma_{2}+\kappa_{3}\gamma_{4}.\nonumber
\end{align}
Now, in order to remove the ladder operators in equation (\ref{Unc1}), we can apply the tilting transformation of the $SU(1,1)$ group with the realization (\ref{su11ab}) as follows
\begin{equation}
D^{\dag}(\xi)H_{I}D(\xi)D^{\dag}(\xi)|\varphi_{1}\rangle =(E^{2}-(\hbar\Delta\omega)^{2})D^{\dag}(\xi)|\varphi_{1}\rangle,
\end{equation}
\begin{equation}
H'_{I}|\varphi_{1}'\rangle=(E^{2}-(\hbar\Delta\omega)^{2})|\varphi'_{1}\rangle.\label{eHsu11}
\end{equation}
Here, $D(\xi)$ is the displacement operator defined in equation (\ref{Dxi}). Notice that in these expressions $H'_{I}=D^{\dag}(\xi)H_{I}D(\xi)$ is the $SU(1,1)$ tilted interaction Hamiltonian and $|\varphi_{1}'\rangle$ its wave function. Thus, by using the tilting transformations (\ref{trsu11ab1})-(\ref{trsu11ab}) of the $Sp(4,R)$ generators the tilted Hamiltonian can be written as
\begin{align}
H'_{I}&=\beta_{-}^{(a)}K_{-}^{(a)}+\beta_{+}^{(a)}K_{+}^{(a)}+\beta_{-}^{(b)}K_{-}^{(b)}+\beta_{+}^{(b)}K_{+}^{(b)}+\beta_{0}^{(ab)}K_{0}^{(ab)}+\beta_{-}^{(ab)}K_{-}^{(ab)}+\beta_{+}^{(ab)}K_{+}^{(ab)}\nonumber\\&+(\alpha_{5}-\alpha_{6})J_{0}+\beta_{-}J_{-}+\beta_{+}J_{+}+\alpha_{11}-\frac{\alpha_{5}+\alpha_{6}}{2},\label{HF1}
\end{align}
where now the $\beta$-coefficients are given by
\begin{align}
&\beta_{-}^{(a)}=\alpha_{1}(\cosh(2|\xi|)+1)+\frac{\alpha_{4}\xi^{*}}{\xi}(\cosh(2|\xi|)-1)-\frac{\alpha_{8}\xi^{*}}{|\xi|}\sinh(2|\xi|),\\&
\beta_{+}^{(a)}=\alpha_{2}(\cosh(2|\xi|)+1)+\frac{\alpha_{3}\xi}{\xi^{*}}(\cosh(2|\xi|)-1)-\frac{\alpha_{9}\xi}{|\xi|}\sinh(2|\xi|),\\&
\beta_{-}^{(b)}=\alpha_{3}(\cosh(2|\xi|)+1)+\frac{\alpha_{2}\xi^{*}}{\xi}(\cosh(2|\xi|)-1)-\frac{\alpha_{9}\xi^{*}}{|\xi|}\sinh(2|\xi|),\\&
\beta_{+}^{(b)}=\alpha_{4}(\cosh(2|\xi|)+1)+\frac{\alpha_{1}\xi}{\xi^{*}}(\cosh(2|\xi|)-1)-\frac{\alpha_{8}\xi}{|\xi|}\sinh(2|\xi|),\\&
\beta_{0}^{(ab)}=(\alpha_{5}+\alpha_{6})\cosh(2|\xi|)+\frac{\alpha_{7}\xi}{|\xi|}\sinh(2|\xi|)+\frac{\alpha_{10}\xi^{*}}{|\xi|}\sinh(2|\xi|),\\&
\beta_{+}^{(ab)}=\frac{(\alpha_{5}+\alpha_{6})\xi}{2|\xi|}\sinh(2|\xi|)+\frac{\alpha_{7}\xi}{2\xi^{*}}(\cosh(2|\xi|)-1)+\frac{\alpha_{10}}{2}(\cosh(2|\xi|)+1),\\&
\beta_{-}^{(ab)}=\frac{(\alpha_{5}+\alpha_{6})\xi^{*}}{2|\xi|}\sinh(2|\xi|)+\frac{\alpha_{7}}{2}(\cosh(2|\xi|)+1)+\frac{\alpha_{10}\xi^{*}}{2\xi}(\cosh(2|\xi|)-1),\\&
\beta_{+}=\alpha_{9}\cosh(2|\xi|)-\frac{\alpha_{3}\xi}{|\xi|}\sinh(2|\xi|)-\frac{\alpha_{2}\xi^{*}}{|\xi|}\sinh(2|\xi|),\\&
\beta_{-}=\alpha_{8}\cosh(2|\xi|)-\frac{\alpha_{4}\xi^{*}}{|\xi|}\sinh(2|\xi|)-\frac{\alpha_{1}\xi}{|\xi|}\sinh(2|\xi|).
\end{align}
If we choose the $\theta$ and $\phi$ parameters of the complex number $\xi=-\frac{\theta}{2}e^{-i\phi}$ as
\begin{equation}
\theta=\tanh^{-1}\left(\frac{\sqrt{(\alpha_{3}\alpha_{8}-\alpha_{1}\alpha_{9})(\alpha_{4}\alpha_{9}-\alpha_{2}\alpha_{8})}}{\alpha_{1}\alpha_{2}-\alpha_{3}\alpha_{4}}\right),\quad\quad\phi=\frac{i}{2}\ln\left[\frac{\alpha_{4}\alpha_{9}-\alpha_{2}\alpha_{8}}{\alpha_{3}\alpha_{8}-\alpha_{1}\alpha_{9}}\right],
\end{equation}
the coefficients $\beta_{+}$ and $\beta_{-}$ vanish, so the tilted Hamiltonian of equation (\ref{HI}) is reduced to
\begin{align}
H'_{I}&=\beta_{-}^{(a)}K_{-}^{(a)}+\beta_{+}^{(a)}K_{+}^{(a)}+\beta_{-}^{(b)}K_{-}^{(b)}+\beta_{+}^{(b)}K_{+}^{(b)}+\beta_{0}^{(ab)}K_{0}^{(ab)}+\beta_{-}^{(ab)}K_{-}^{(ab)}+\beta_{+}^{(ab)}K_{+}^{(ab)}\nonumber\\&+(\alpha_{5}-\alpha_{6})J_{0}+\alpha_{11}-\frac{\alpha_{5}+\alpha_{6}}{2}.\label{Hsu11}
\end{align}
It is important to note that the parameters $\theta$ and $\phi$ are dimensionless as we expected. Now, we apply the tilting transformation to the eigenvalue equation (\ref{eHsu11}) in terms of the realization (\ref{su2}) of the $SU(2)$ group. For this case, we define the new tilting Hamiltonian $H''_{I}=D^{\dag}(\chi)H'_{I}D(\chi)$ and its wave function $|\varphi_{1}''\rangle=D^{\dag}(\chi)|\varphi_{1}'\rangle$. Therefore, we find that the $SU(2)$ tilted Hamiltonian $H''_{I}$ results to be
\begin{align}
H''_{I}&=\alpha_{-}^{(a)}K_{-}^{(a)}+\alpha_{+}^{(a)}K_{+}^{(a)}+\alpha_{-}^{(b)}K_{-}^{(b)}+\alpha_{+}^{(b)}K_{+}^{(b)}+\beta_{0}^{(ab)}K_{0}^{(ab)}+\alpha_{-}^{(ab)}K_{-}^{(ab)}+\alpha_{+}^{(ab)}K_{+}^{(ab)}\nonumber\\&+(\alpha_{5}-\alpha_{6})D^{\dag}(\chi)J_{0}D(\chi)+\alpha_{11}-\frac{\alpha_{5}+\alpha_{6}}{2},\label{Hsu11su2}
\end{align}
where the new $\alpha$-coefficients are given by
\begin{align}
&\alpha_{-}^{(a)}=\frac{\beta_{-}^{(a)}}{2}(\cos(2|\chi|)+1)-\frac{\beta_{-}^{(b)}\chi^{*}}{2\chi}(\cos(2|\chi|)-1)+\frac{\beta_{-}^{(ab)}\chi^{*}}{|\chi|}\sin(2|\chi|),\\&
\alpha_{+}^{(a)}=\frac{\beta_{+}^{(a)}}{2}(\cos(2|\chi|)+1)-\frac{\beta_{+}^{(b)}\chi}{2\chi^{*}}(\cos(2|\chi|)-1)+\frac{\beta_{+}^{(ab)}\chi}{|\chi|}\sin(2|\chi|),\\&
\alpha_{-}^{(b)}=\frac{\beta_{-}^{(b)}}{2}(\cos(2|\chi|)+1)-\frac{\beta_{-}^{(a)}\chi}{2\chi^{*}}(\cos(2|\chi|)-1)-\frac{\beta_{-}^{(ab)}\chi}{|\chi|}\sin(2|\chi|),\\&
\alpha_{+}^{(b)}=\frac{\beta_{+}^{(b)}}{2}(\cos(2|\chi|)+1)-\frac{\beta_{+}^{(a)}\chi^{*}}{2\chi}(\cos(2|\chi|)-1)-\frac{\beta_{+}^{(ab)}\chi^{*}}{|\chi|}\sin(2|\chi|),\\&
\alpha_{+}^{(ab)}=\beta_{+}^{(ab)}\cos(2|\chi|)-\frac{\beta_{+}^{(a)}\chi^{*}}{2|\chi|}\sin(2|\chi)+\frac{\beta_{+}^{(b)}\chi}{2|\chi|}\sin(2|\chi|),\\&
\alpha_{-}^{(ab)}=\beta_{-}^{(ab)}\cos(2|\chi|)-\frac{\beta_{-}^{(a)}\chi}{2|\chi|}\sin(2|\chi|)+\frac{\beta_{-}^{(b)}\chi^{*}}{2|\chi|}\sin(2|\chi|).
\end{align}
This transformation will allow us to remove the ladder operators $K_{\pm}^{(ab)}$ of the tilting Hamiltonian $H''_{I}$ by choosing suitably the parameters $\theta$ and $\phi$ of the complex number $\chi=-\frac{\theta}{2}e^{-i\phi}$ that make the coefficients $\alpha_{-}^{(ab)}$ and $\alpha_{+}^{(ab)}$ zero. These values of $\theta$ and $\phi$ are given by
\begin{equation}
\theta=\tan^{-1}\left[2\frac{\sqrt{(\beta_{-}^{(a)}\beta_{+}^{(ab)}+\beta_{+}^{(b)}\beta_{-}^{(ab)})(\beta_{-}^{(b)}\beta_{+}^{(ab)}+\beta_{+}^{(a)}\beta_{-}^{(ab)})}}{\beta_{-}^{(b)}\beta_{+}^{(b)}-\beta_{-}^{(a)}\beta_{+}^{(a)}}\right],\quad\quad\phi=i\ln\left[\sqrt{\frac{\beta_{-}^{(ab)}\beta_{+}^{(ab)}+\beta_{+}^{(a)}\beta_{-}^{(ab)}}{\beta_{-}^{(a)}+\beta_{+}^{(ab)}+\beta_{+}^{(b)}\beta_{-}^{(ab)}}}\right].
\end{equation}
Therefore, the tilting Hamiltonian $H''_{I}$ (\ref{Hsu11su2}) is reduced to
\begin{align}
H''_{I}&=\alpha_{-}^{(a)}K_{-}^{(a)}+\alpha_{+}^{(a)}K_{+}^{(a)}+\alpha_{-}^{(b)}K_{-}^{(b)}+\alpha_{+}^{(b)}K_{+}^{(b)}+\beta_{0}^{(ab)}K_{0}^{(ab)}+\alpha_{11}-\frac{\alpha_{5}+\alpha_{6}}{2}\nonumber\\&+(\alpha_{5}-\alpha_{6})D^{\dag}(\chi)J_{0}D(\chi).\label{FHsu11su2}
\end{align}

Following the previous procedure, we apply the tilting transform of equation (\ref{su11xsu11}) to the Hamiltonian $H''_{I}$. If we consider that the boson operators $\hat{a}$ and $\hat{b}$ commute, we can transform the Hamiltonian $H''_{I}$ as
\begin{align}
H_{I}'''=&\sqrt{{\beta_{0}^{(ab)}}^{2}-4\alpha_{+}^{(a)}\alpha_{-}^{(a)}}K_{0}^{(a)}+\sqrt{{\beta_{0}^{(ab)}}^{2}-4\alpha_{+}^{(b)}\alpha_{-}^{(b)}}K_{0}^{(b)}+(\alpha_{5}-\alpha_{6})D(\xi)_{a,b}^{\dag}D^{\dag}(\chi)J_{0}D(\chi)D(\xi)_{a,b}\nonumber\\&+\alpha_{11}-\frac{\alpha_{5}+\alpha_{6}}{2}.\label{H'''}
\end{align}
Here, we have removed the ladder operators $K_{\pm}^{(a)}$ and $K_{\pm}^{b)}$ (see Appendix A) by choosing the parameters $\theta_{a}$, $\phi_{a}$ and $\theta_{b}$, $\phi_{b}$ of the complex numbers $\xi_{a}=\frac{\theta_{a}}{2}e^{-i\phi_{a}}$ and $\xi_{b}=\frac{\theta_{b}}{2}e^{-i\phi_{b}}$ as
\begin{align}
&\theta_{a}=\tanh^{-1}\left[\frac{2\sqrt{\alpha_{+}^{(a)}\alpha_{-}^{(a)}}}{\beta_{0}^{(ab)}}\right],\quad\quad\phi_{a}=i\ln\left[\frac{\beta_{o}^{(ab)}\alpha}{2\alpha_{-}^{(a)}(2\beta+1)}\right],\\&
 \theta_{b}=\tanh^{-1}\left[\frac{2\sqrt{\alpha_{+}^{(b)}\alpha_{-}^{(b)}}}{\beta_{0}^{(ab)}}\right],\quad\quad\phi_{a}=i\ln\left[\frac{\beta_{o}^{(ab)}\alpha}{2\alpha_{-}^{(b)}(2\beta+1)}\right].
\end{align}
It is necessary to point out that the term $D(\xi)_{a,b}^{\dag}D^{\dag}(\chi)J_{0}D(\chi)D(\xi)_{a,b}$ could return the ladder operators we have eliminated in each transformation previously applied, but this will depend on the value of the constants $\kappa_{i}$ and $\gamma_{i}$ which we choose from the Hamiltonian (\ref{HI}). Thus, to be able to give a solution of the eigenvalue equation $H'''\varphi_{1'''}=E^{2}-((\hbar\Delta)^{2})\varphi_{1'''}$ of the Hamiltonian (\ref{H'''}) we are going to consider that $\alpha_{5}=\alpha_{6}$, namely
\begin{equation}
\kappa_1\gamma_2+\kappa_2\gamma_1=\kappa_4\gamma_3+\kappa_3\gamma_4.\label{cond}
\end{equation}
With this assumption, the Hamiltonian $H_{I}'''$ is reduced to the diagonal form
\begin{equation}
H_{I}'''=\sqrt{{\beta_{0}^{(ab)}}^{2}-4\alpha_{+}^{(a)}\alpha_{-}^{(a)}}K_{0}^{(a)}+\sqrt{{\beta_{0}^{(ab)}}^{2}-4\alpha_{+}^{(b)}\alpha_{-}^{(b)}}K_{0}^{(b)}+\alpha_{11}-\alpha_{5}.\label{Hf}
\end{equation}

Now, let us now look the eigenfunctions $|\varphi_1'''\rangle= D(\xi)_{a,b}^{\dag}|\varphi_{1}''\rangle$ of the Hamiltonian $H'''$. Since the operator $K_{0}$ is the Hamiltonian of the two-dimensional harmonic oscillator and commutes with $J_{0}$, we have that the eigenfunctions of $H'''$ are given by
\begin{equation}
\varphi'''_{n_{l},m_n}(\rho,\phi)=\frac{1}{\sqrt{\pi}}e^{im_n\phi}(-1)^{n_{l}}\sqrt{\frac{2(n_{l})!}{(n_{l}+m_n)!}}\rho^{m_n}L^{m_n}_{n_{l}}(\rho^{2})e^{-1/2\rho^{2}}\label{Poly1},
\end{equation}
where $n_l$ is the left chiral quantum number. The eigenfunctions of the Hamiltonian of a two-level system (\ref{H}) under the condition of equation (\ref{cond}) are obtained from the relationship
\begin{equation}
|\varphi_{1}\rangle=D(\xi)D(\chi)D(\xi)_{a,b}|\varphi'''_{1}\rangle.
\end{equation}

By considering the action of the operators $\hat{a},\hat{a}^{\dag}$ and $\hat{b},\hat{b}^{\dag}$ on the basis $|n,m_n\rangle$, we have
\begin{equation}
K_{0}|n,m_n\rangle=\frac{1}{2}(\hat{a}^{\dag}\hat{a}+\hat{b}^{\dag}\hat{b}+1)|n,m_n\rangle=\frac{n+1}{2}|n,m_n\rangle,\nonumber
\end{equation}
\begin{equation}
J_{0}|n,m_n\rangle=\frac{1}{2}(\hat{a}^{\dag}\hat{a}-\hat{b}^{\dag}\hat{b})|n,m_n\rangle=\frac{m_{n}}{2}|n,m_n\rangle.
\end{equation}
Thus, from these results we can obtain that the energy spectrum of the general interaction Hamiltonian
of a two-level system (\ref{HI}) (submitted to condition of equation (\ref{cond})) is given by
\begin{align}
E_{n,m_n}&=\pm\left[\left(\frac{\hbar\omega_{0}}{2}-\frac{\hbar\omega_{1}}{2}-\frac{\hbar\omega_{2}}{2}\right)^{2}+\left(\sqrt{{\beta_{0}^{(ab)}}^{2}-4\alpha_{+}^{(a)}\alpha_{-}^{(a)}}+\sqrt{{\beta_{0}^{(ab)}}^{2}-4\alpha_{+}^{(b)}\alpha_{-}^{(b)}}\right)\frac{n+1}{4}\nonumber\right.\\&\left.+\left(\sqrt{{\beta_{0}^{(ab)}}^{2}-4\alpha_{+}^{(a)}\alpha_{-}^{(a)}}-\sqrt{{\beta_{0}^{(ab)}}^{2}-4\alpha_{+}^{(b)}\alpha_{-}^{(b)}}\right)\frac{m_{n}}{4}+\alpha_{11}-\alpha_{5}\right]^{\frac{1}{2}}.\label{energy}
\end{align}
Analogously, if we apply the same procedure to the uncoupled equation for the other spinor component $|\varphi _{2}\rangle$ we can obtain a similar expression of the energy spectrum but with a shift in energy levels. However, since both energies of the spinor components belong to the same solution, we must choose suitably the quantum numbers of the spinor components so that their energy spectrum matches.

Therefore, we have obtained the energy eigenvalues of the most general interaction Hamiltonian of a two-level system in two dimensional geometry by using of the $sp(4,R)$ Lie algebra, which is based on bosonic annihilation $\hat{a}$, $\hat{b}$ and creation $\hat{a}^{\dag}$, $\hat{b}^{\dag}$ operators. It is worthwhile to mention that to get this result we have imposed an order on the application of the transformations. In our work, we first used the displacement operator of the $SU(1,1)$ group based on the two-mode boson realization (\ref{su11ab}). Then, we have used the displacement operator of the $SU(2)$ group based on realization (\ref{su2}), and finally we have used the displacement operators of the $SU(1,1)$ group based on the single-mode boson realizations (\ref{su11a}) and (\ref{su11b}).

\section{Special cases of the general Hamiltonian}

In this Section, we will study some particular cases of the general Hamiltonian of equation (\ref{H}). These cases are of physical interest and are obtained by giving specific values to the parameters $\kappa_{i}$ and $ \gamma_{i}$. The Hamiltonians here presented have an exact solution and can be written in terms of an appropriate Lie algebra.

\subsection{The Jaynes-Cummings model}

The Jaynes Cummings model (sometimes abbreviated JCM) is a theoretical model in quantum optics which describes the system of a two-level atom interacting with a quantized mode of an optical cavity (or a bosonic field). The Hamiltonian of this model can be obtained by setting $\kappa_{2}=\kappa_{3}=\kappa_{4}=\gamma_{1}=\gamma_{3}=\gamma_{4}=0$ and $\kappa_{1}=\gamma_{2}=\kappa$ \cite{Koc}. Hence, with these values of the parameters $\kappa_{i}$ and $\gamma_{i}$ the Hamiltonian of equation (\ref{H}) is reduced to
\begin{equation}
\hat{H}_{JC}=\hbar\omega_{1}\hat{a}^{\dag}\hat{a}+\frac{\hbar\omega_{0}}{2}\sigma_{0}+\kappa(\hat{a}\sigma_{+}+\hat{a}^{\dag}\sigma_{-}).
\end{equation}
In this case, the Hamiltonian can be decomposed as $\hat{H}_{JC}=\hat{H}_{0}+\hat{H}_{JC_{I}}$, where
\begin{equation}
H_{0}=\hbar\omega_{1}\left(\hat{a}^{\dag} \hat{a}+\frac{\sigma_{0}}{2}\right),
\end{equation}
and with the definition $\Delta\omega=\omega_{0}-\omega_{1}$, the interaction Hamiltonian (\ref{HI}) takes the form
\begin{equation}
H_{JC_{I}}=\frac{\Delta\omega}{2}\sigma_{0}+\kappa(\hat{a}\sigma_{+}+\hat{a}^{\dag}\sigma_{-}).
\end{equation}
The uncoupled equation (\ref{Unc1}) for the spinor component $|\varphi_1\rangle$ of the JC Hamiltonian is
\begin{equation}
\left[2\kappa^{2}K_{0}^{(a)}+\frac{\kappa^{2}}{2}\right]|\varphi_1\rangle=\left(E^{2}-\frac{(\Delta\omega\hbar)^{2}}{4}\right)|\varphi_1\rangle.
\end{equation}
From the energy spectrum of equation (\ref{energy}), we obtain that the eigenvalues of the interaction JC Hamiltonian are given by \cite{Koc}
\begin{equation}
E=\pm\frac{1}{2}\sqrt{(\Delta\omega\hbar)^{2}+4\kappa^{2}(n+1)}.
\end{equation}
It is important to note that since the interaction Hamiltonian commutes with $H_{0}$, the energy spectrum of the Jaynes-Cummings Hamiltonian is
\begin{equation}
E_{JC}=\hbar\omega_{1}\left(n+\frac{1}{2}\right)+\pm\frac{1}{2}\sqrt{(\Delta\omega\hbar)^{2}+4\kappa^{2}(n+1)}.
\end{equation}

\subsection{$2+1$ Dirac-Moshinsky oscillator}

The Dirac oscillator is a relativistic problem such that its non-relativistic limit leads to the Schr\"odinger equation of the harmonic oscillator. The Hamiltonian of the Dirac-Moshinsky oscillator in 2+1 dimensions can be obtained by setting the parameters $\omega_{1}=\omega_{2}=\kappa_{2}=\kappa_{3}=\kappa_{4}=\gamma_{1}=\gamma_{3}=\gamma_{4}=0$, $\kappa_{1}=\gamma_{2}=2ic\sqrt{m\omega\hbar}$, and $\omega_{0}=mc^{2}/\hbar$ . With these definitions the Hamiltonian of equation (\ref{H}) is reduced to \cite{Koc,Bermudez}
\begin{equation}
H=mc^{2}\sigma_{0}+2ic\sqrt{m\omega\hbar}(\hat{a}\sigma_{+}+\hat{a}^{\dag}\sigma_{-}).
\end{equation}
Here, the algebraic form of the uncoupled equation (\ref{Unc1}) for the spinor component $|\varphi_1\rangle$ is given by
\begin{equation}
-4c^{2}m\omega\hbar\left[2K_{0}^{(a)}+\frac{1}{2}\right]|\varphi_1\rangle=\left(E^{2}-m^{2}c^{4}\right)|\varphi_1\rangle.
\end{equation}
Then, from equation (\ref{energy}) we obtain the following energy spectrum for the Dirac-Moshinsky oscillator in 2+1 dimensions\cite{Koc}
\begin{equation}
E=\pm\sqrt{m^{2}c^{4}-4\hbar\omega mc^{2}(n+1)}.
\end{equation}

\subsection{The generalized Jaynes-Cummings model}

Another interesting Hamiltonian with one oscillation mode is a generalization of the Jaynes-Cummings model, formed by a combination of a Jaynes-Cummings and an Anti-Jaynes-Cummings model \cite{Nos}. This Hamiltonian is obtained from equation (\ref{H}) by setting the parameters $\kappa_{3}=\kappa_{4}=\gamma_{3}=\gamma_{4}=0$, $\kappa_{1}=\hbar f^{*}$, $\gamma_{1}=\hbar g*$, $\kappa_{2}=\hbar g$, $\gamma_{2}=\hbar f$ and the frequencies $\omega_{1}= \omega_{2}=0$ and $\omega_{0}=\frac{mc^{2}}{\hbar}$. From this election of the parameters, the Hamiltonian (\ref{H}) is reduced to
\begin{equation}
H=\hbar\left[\sigma_-(g^*a+f a^{\dag})+\sigma_+(ga^{\dag}+f^*a)\right]+mc^2\sigma_z.
\end{equation}
For this model, the uncoupled equation (\ref{Unc1}) for the spinor component $|\varphi_1\rangle$ is in terms of the
the $SU(1,1)$ Lie algebra realization of equation (\ref{su11a}) as
\begin{equation}
\left[\alpha_{5}K_0+2\alpha_{2}K_{+}+2\alpha_{1}K_{-}+\alpha_{11}-\alpha_{5}\right]|\varphi_1\rangle=(E^2-m^2c^4)|\varphi_1\rangle\label{unc1},
\end{equation}
where $\alpha_{11}=\hbar^{2}|f|^{2}$, $\alpha_{5}=2\hbar^2(|g|^2+|f|^2)$, $\alpha_{1}=\hbar^{2}f^{*}g^{*}$ and $\alpha_{2}=\hbar^{2}gf$.

From the $\alpha's$ values we have that $\beta_{0}^{(ab)}=\alpha_{5}$, $\alpha_{+}^{(a)}=2\alpha_2$ and $\alpha_{-}^{(a)}=2\alpha_1$. Therefore, from the $\beta's$ values and expression (\ref{Hf}) we can obtain the energy spectrum for this model \cite{Nos}
\begin{equation}
E=\pm\sqrt{\hbar^2(|g|^2-|f|^2)n+m^2c^4}.\label{spectrum}
\end{equation}

\subsection{Single two-level atom interacting with two quantized modes}

The Hamiltonian of a single two-level atom interacting with two quantized modes can be obtained from the modification of the JC Hamiltonian such that this model is the addition of two Jaynes-Cummings models with different modes of oscillation. This Hamiltonian is often called the Modified Jaynes-Cummings model (MJC) and can be obtained by choosing the parameters $\kappa_{2}=\kappa_{4}=\gamma_{1}=\gamma_{3}=0$, $\kappa_{1}=\gamma_{2}=\lambda_{1}$ and $\kappa_{3}=\gamma_{4}=\lambda_{2}$. Then, the Hamiltonian (\ref{H}) becomes \cite{Koc}
\begin{equation}
H_{MJC}=\hbar\omega\hat{a}^{\dag}\hat{a}+\hbar\omega\hat{b}^{\dag}\hat{b}+\frac{\hbar\omega_{0}}{2}\sigma_{0}+(\lambda_{1}\hat{a}+\lambda_{2}\hat{b})\sigma_{+}+(\lambda_{1}\hat{a}^{\dag}+\lambda_{2}\hat{b}^{\dag})\sigma_{-}.
\end{equation}
In this case, the Hamiltonian can be decomposed as $\hat{H}_{MJC}=\hat{H}_{0}+\hat{H}_{MJC_{I}}$, where
\begin{equation}
H_{0}=\hbar\omega_{1}\left(\hat{a}^{\dag} \hat{a}+\frac{\sigma_{0}}{2}\right)+\hbar\omega_{2}\left(\hat{b}^{\dag} \hat{b}+\frac{\sigma_{0}}{2}\right),
\end{equation}
and the interaction Hamiltonian of the MJC model is given by
\begin{equation}
H_{MJC_{I}}=\hbar\left(\frac{\omega_{0}-\omega_{1}-\omega_{2}}{2}\right)\sigma_{0}+(\lambda_{1}\hat{a}+\lambda_{2}\hat{b})\sigma_{+}+(\lambda_{1}\hat{a}^{\dag}+\lambda_{2}\hat{b}^{\dag})\sigma_{-}.
\end{equation}
Hence, the $\alpha$-constants take the values $\alpha_{1}=\alpha_{2}=\alpha_{3}=\alpha_{4}=\alpha_{7}=\alpha_{10}=0$, $\alpha_{5}=\lambda_{1}^{2}$, $\alpha_{6}=\lambda_{2}^{2}$, $\alpha_{8}=\alpha_{9}=\lambda_{1}\lambda_{2}$ and $\alpha_{11}=\lambda_{1}^{2}+\lambda_{2}^{2}$. The uncoupled equation (\ref{Unc1}) for the spinor component $|\varphi_1\rangle$ takes the form
\begin{align}
H_{MJC_{I}}|\varphi_{1}\rangle &=\left[(\lambda_{1}^{2}-\lambda_{2}^{2})J_{0}+\lambda_{1}\lambda_{2}J_{+}+\lambda_{1}\lambda_{2}J_{-}+(\lambda_{1}^{2}+\lambda_{2}^{2})K_{0}^{(ab)}+\frac{(\lambda_{1}^{2}+\lambda_{2}^{2})}{2}\right]|\varphi_{1}\rangle\nonumber\\&=\left(E^{2}-\frac{(\hbar\Delta\omega)^{2}}{4}\right)|\varphi_{1}\rangle.
\end{align}
Whit these values of the $\alpha$ constants we found that $\beta_{0}^{(ab)}=\lambda_{1}^{2}+\lambda_{2}^{2}$,  $\beta_{+}=\beta_{-}=\lambda_{1}\lambda_{2}$. Moreover, the above expression is a equation of the type (\ref{AHsu2}) of Appendix B. By using the tilting transformation with the $SU(2)$ displacement operator (see equation (\ref{ADHsu2})) we obtain
\begin{equation}
\left[\sqrt{(\lambda_{1}^{2}-\lambda_{2}^{2})^{2}+4\lambda_{1}^{2}\lambda_{2}^{2}}J_{0}+(\lambda_{1}^{2}+\lambda_{2}^{2})K_{0}^{(ab)}+\frac{(\lambda_{1}^{2}+\lambda_{2}^{2})}{2}\right]|\varphi'_{1}\rangle=\left(E^{2}-\frac{(\hbar\Delta\omega)^{2}}{4}\right)|\varphi'_{1}\rangle.
\end{equation}
 Therefore, since $H_{0}$ commutes with $H_{MJC_{I}}$ the energy spectrum of the Modified Jaynes-Cummings Hamiltonian is
\begin{equation}
E=\hbar\left(\frac{\omega_1+\omega_2}{2}\right)\left(n+1\right)+\hbar\left(\frac{\omega_1-\omega_2}{2}\right)m\pm\frac{1}{2}\sqrt{(\hbar\Delta\omega)^{2}+4\left(\lambda_{1}^{2}+\lambda_{2}^{2}\right)\left(\frac{n}{2}+\frac{m}{2}+1\right)}.
\end{equation}

\subsection{The two-mode Jaynes-Cummings-Anti-Jaynes-Cummings model}

This model is a linear combination of the Jaynes-Cummings and Anti-Jaynes-Cummings models with different modes of oscillation. To obtain this Hamiltonian we have to choose the parameters as follows $\kappa_{2}=\kappa_{3}=\gamma_{1}=\gamma_{4}=0$, $\kappa_{1}=\gamma_{2}=\lambda_{1}$ and $\kappa_{4}=\gamma_{3}=\lambda_{2}$. Therefore, the Hamiltonian (\ref{H}) becomes \cite{Nos1}
\begin{equation}
H_{JC-AJC}=\hbar\omega\hat{a}^{\dag}\hat{a}+\hbar\omega\hat{b}^{\dag}\hat{b}+\frac{\hbar\omega_{0}}{2}\sigma_{0}+(\lambda_{1}\hat{a}+\lambda_{2}\hat{b}^{\dag})\sigma_{+}+(\lambda_{1}\hat{a}^{\dag}+\lambda_{2}\hat{b})\sigma_{-}.
\end{equation}
We can split this Hamiltonian as $H_{JC-AJC}=H_{0}+H_{JC-AJC_{I}}$, where
\begin{equation}
H_{0}=\hbar\omega_{1}\left(\hat{a}^{\dag} \hat{a}+\frac{\sigma_{0}}{2}\right)+\hbar\omega_{2}\left(\hat{b}^{\dag} \hat{b}-\frac{\sigma_{0}}{2}\right),
\end{equation}
and the interaction Hamiltonian of the JC-AJC model is given by the following expression
\begin{equation}
H_{JC-AJC_{I}}=\hbar\left(\frac{\omega_{0}+\omega_{2}-\omega_{1}}{2}\right)\sigma_{0}+(\lambda_{1}\hat{a}+\lambda_{2}\hat{b}^{\dag})\sigma_{+}+(\lambda_{1}\hat{a}^{\dag}+\lambda_{2}\hat{b})\sigma_{-}.
\end{equation}
In this problem the $\alpha$-constants take the values $\alpha_{1}=\alpha_{2}=\alpha_{3}=\alpha_{4}=\alpha_{8}=\alpha_{9}=0$, $\alpha_{5}=\lambda_{1}^{2}$, $\alpha_{6}=\lambda_{2}^{2}$, $\alpha_{7}=\lambda_{1}\lambda_{2}$, $\lambda_{10}=\lambda_{1}\lambda_{2}$ and $\lambda_{11}=\lambda_{1}^{2}$. Thus, the uncoupled equation (\ref{Unc1}) for the upper spinor component $|\varphi_1\rangle$ takes the form
\begin{align}
H_{JC-AJC_{I}}|\varphi_{1}\rangle &=\left[(\lambda_{1}^{2}+\lambda_{2}^{2})K_{0}^{(ab)}+\lambda_{1}\lambda_{2}K_{+}^{(ab)}+\lambda_{1}\lambda_{2}K_{-}^{(ab)}+(\lambda_{1}^{2}-\lambda_{2}^{2})J_{0}+\frac{(\lambda_{1}^{2}-\lambda_{2}^{2})}{2}\right]|\varphi_{1}\rangle\nonumber\\&=\left(E^{2}-\frac{(\hbar\Delta\omega)^{2}}{4}\right)|\varphi_{1}\rangle.\label{MJC}
\end{align}
The constants $\beta_{0}^{(ab)}$, $\beta_{+}^{(ab)}$ and $\beta_{-}$ of the expression (\ref{HF1}) have the values
\begin{equation}
\beta_{0}^{(ab)}=(2\beta+1)(\lambda_{1}^{2}+\lambda_{2}^{2})+\frac{\xi^{*}}{|\xi|}\alpha\lambda_{1}\lambda_{2}+\frac{\xi}{|\xi|}\alpha\lambda_{1}\lambda_{2},
\end{equation}
\begin{equation}
\beta_{+}^{(ab)}\frac{\xi}{2|\xi|}\alpha(\lambda_{1}^{2}+\lambda_{2}^{2})+(\beta+1)\lambda_{1}\lambda_{2}+\frac{\xi}{\xi^{*}}\beta\lambda_{1}\lambda_{2},
\end{equation}
\begin{equation}
\beta_{-}^{(ab)}\frac{\xi^{*}}{2|\xi|}\alpha(\lambda_{1}^{2}+\lambda_{2}^{2})+\frac{\xi^{*}}{\xi}\beta\lambda_{1}\lambda_{2}+(\beta+1)\lambda_{1}\lambda_{2}.
\end{equation}
The expression (\ref{MJC}) is an equation of the type (\ref{AHsu11}) of Appendix A (see equation (\ref{ADHsu11})) and after the $SU(1,1)$ tilting transformation we obtain the following result
\begin{equation}
\left[\sqrt{(\lambda_{1}^{2}+\lambda_{2}^{2})^{2}-4\lambda_{1}^{2}\lambda_{2}^{2}}K_{0}^{(ab)}+(\lambda_{1}^{2}-\lambda_{2}^{2})J_{0}+\frac{(\lambda_{1}^{2}-\lambda_{2}^{2})}{2}\right]|\varphi'_{1}\rangle=\left(E^{2}-\frac{\hbar^{2}\omega^{2}}{4}\right)|\varphi'_{1}\rangle.
\end{equation}
Therefore, since $H_{0}$ commutes with $H_{JC-AJC_{I}}$ the energy spectrum of the JC-AJC Hamiltonian is now given by \cite{Nos1}
\begin{equation}
E=\hbar\left(\frac{\omega_1+\omega_2}{2}\right)n+\hbar\left(\frac{\omega_1-\omega_2}{2}\right)\left(m+1\right)\pm\frac{1}{2}\sqrt{(\hbar\Delta\omega)^{2}+4\left(\lambda_{1}^{2}-\lambda_{2}^{2}\right)\left(\frac{n}{2}+\frac{m}{2}+1\right)}.
\end{equation}

\section{Discussion and Conclusion}

In this paper we first review the main properties of the $SU(1,1)$ and $SU(2)$ group bosonic realizations. With these properties we constructed the generators of the $Sp(4,R)$ group and the commutation relations satisfied by these operators were obtained. With this theory, we computed the similarity transformation of the $Sp(4,R)$ generators in terms of the $SU(1,1)$ and $SU(2)$ displacement operators.

Our procedure allowed to introduce a novel algebraic method to solve exactly the interaction part of the most general Hamiltonian of a two-level system in two-dimensional geometry. More explicitly, to obtain the exact energy spectrum of our problem we applied three tilting transformations to diagonalize the general Hamiltonian. The importance of studying this Hamiltonian is due to the fact that, as particular cases, it reduces to other models of vital importance in quantum mechanics and in quantum optics, as the Jaynes-Cummings model, the Modified Jaynes-Cummings model, the Jaynes-Cummings model with one and two modes, and the $2+1$ Dirac-Moshinsky oscillator, among others. In Section 4 of our paper, we reproduced the energy spectrum of these problems.

Is is important to note that this general Hamiltonian was previously studied in reference \cite{Koc}. In their work, the authors introduced an alternative and completely different algebraic method to solve some particular cases of the Hamiltonian.

\section*{Acknowledgments}

This work was partially supported by SNI-M\'exico, EDI-IPN, SIP-IPN project number $20195316$. The authors would like to thank Professor Roberto Daniel Mota Esteves and Professor V\'ictor David Granados Garc\'ia for their helpful comments and suggestions made to our work.

\renewcommand{\theequation}{A.\arabic{equation}}
\setcounter{equation}{0}

\section*{Appendix A. The $SU(1,1)$ group theory and its tilting transformation}

The $su(1,1)$ Lie algebra is defined by the following commutation relations \cite{Vourdas}
\begin{eqnarray}
[K_{0},K_{\pm}]=\pm K_{\pm},\quad\quad [K_{-},K_{+}]=2K_{0}.\label{comm}
\end{eqnarray}
The theory of unitary irreducible representations of the $su(1,1)$ Lie algebra has been
studied in several works \cite{ADAMS} and it is based on equations (\ref{k+n})-(\ref{Cas}).
Thus, a representation of $su(1,1)$ algebra is determined by the number $k$, called the Bargmann index.

The displacement operator $D(\xi)$ for this algebra is defined in terms of the creation and annihilation operators $K_+, K_-$ as
\begin{equation}
D(\xi)=\exp(\xi K_{+}-\xi^{*}K_{-}),\label{do}
\end{equation}
where $\xi=-\frac{1}{2}\tau e^{-i\varphi}$, $-\infty<\tau<\infty$ and $0\leq\varphi\leq2\pi$. Since the ladder operators $K_{\pm}$ satisfy the properties $K^{\dag}_{+}=K_{-}$ and $K^{\dag}_{-}=K_{+}$, it can be shown that the displacement operator possesses the property
\begin{equation}
D^{\dag}(\xi)=\exp(\xi^{*} K_{-}-\xi K_{+})=D(-\xi).
\end{equation}

A more useful representation of the displacement operator $D(\xi)$ is given by the so called normal form of this operator
\begin{equation}
D(\xi)=\exp(\zeta K_{+})\exp(\eta K_{3})\exp(-\zeta^*
K_{-})\label{normal},
\end{equation}
where $\xi=-\frac{1}{2}\tau e^{-i\varphi}$, $\zeta=-\tanh
(\frac{1}{2}\tau)e^{-i\varphi}$ and $\eta=-2\ln \cosh
|\xi|=\ln(1-|\zeta|^2)$ \cite{GER}.

The $SU(1,1)$ Perelomov coherent state is defined as the action of the displacement operator $D(\xi)$
on the lowest normalized state $|k,0\rangle$. The normal form of the displacement operator and equations (\ref{k+n})-(\ref{k0n}) can be used to obtain the following expression of the Perelomov coherent states \cite{PERL}
\begin{equation}
|\zeta\rangle=D(\xi)|k,0\rangle=(1-|\xi|^2)^k\sum_{s=0}^\infty\sqrt{\frac{\Gamma(n+2k)}{s!\Gamma(2k)}}\xi^s|k,s\rangle.\label{PCN}
\end{equation}
The $SU(1,1)$ Perelomov number coherent state $|\zeta,k,n\rangle$ is defined as the action of the displacement operator $D(\xi)$ on an arbitrary
excited state $|k,n\rangle$ \cite{Nos1}
\begin{eqnarray}
|\zeta,k,n\rangle &=&\sum_{s=0}^\infty\frac{\zeta^s}{s!}\sum_{j=0}^n\frac{(-\zeta^*)^j}{j!}e^{\eta(k+n-j)}
\frac{\sqrt{\Gamma(2k+n)\Gamma(2k+n-j+s)}}{\Gamma(2k+n-j)}\nonumber\\
&&\times\frac{\sqrt{\Gamma(n+1)\Gamma(n-j+s+1)}}{\Gamma(n-j+1)}|k,n-j+s\rangle.\label{PNCS}
\end{eqnarray}

The tilting transformation of the $su(1,1)$ Lie algebra generators are computed by using the displacement operator $D(\xi)$ and the Baker-Campbell-Hausdorff identity
\begin{equation*}
e^{A}Be^{-A}=B+[B,A]+\frac{1}{2!}[[B,A],A]+\frac{1}{3!}[[[B,A]A]A]+...
\end{equation*}
These results are give by
\begin{equation}
D^{\dag}(\xi)K_{+}D(\xi)=\frac{\xi^{*}}{|\xi|}\alpha K_{0}+\beta\left(K_{+}+\frac{\xi^{*}}{\xi}K_{-}\right)+K_{+},\label{DK+}
\end{equation}
\begin{equation}
D^{\dag}(\xi)K_{-}D(\xi)=\frac{\xi}{|\xi|}\alpha K_{0}+\beta\left(K_{-}+\frac{\xi}{\xi^{*}}K_{+}\right)+K_{-},\label{DK-}
\end{equation}
\begin{equation}
D^{\dag}(\xi)K_{0}D(\xi)=(2\beta+1)K_{0}+\frac{\alpha\xi}{2|\xi|}K_{+}+\frac{\alpha\xi^{*}}{2|\xi|}K_{-},\label{DK0}
\end{equation}
where $\alpha=\sinh(2|\xi|)$ and $\beta=\frac{1}{2}\left[\cosh(2|\xi|)-1\right]$.

A particular realization of the $su(1,1)$ Lie algebra is given by the Jordan-Schwinger operators
\begin{equation}
K_0=\frac{1}{2}\left(a^{\dag}a+b^{\dag}b+1\right), \quad K_+=a^{\dag}b^{\dag},\  \quad K_-= ba,\label{gen11}
\end{equation}
where the two sets of operators $(a,a^{\dag})$ and $(b,b^{\dag})$ satisfy the bosonic algebra
\begin{equation}
[a,a^{\dag}]=[b,b^{\dag}]=1, \quad\quad[a,b^{\dag}]=[a,b]=0.
\end{equation}
If we introduce the difference of the number operators of the two oscillators $N_d$, it can be shown that it commutes with all
the generators of this algebra
\begin{equation}
[N_d,K_0]=[N_d,K_+]=[N_d,K_-]=0.
\end{equation}
Explicitly, $N_d$ and the Casimir operator for this realization are give by \cite{vourdasanalytic}
\begin{equation}
K^2=\frac{1}{4}N_d^2-\frac{1}{4}, \quad\quad N_d=b^{\dag}b-a^{\dag}a.
\end{equation}

Now, as an application of the theory developed above we are going to solve the following eigenvalue problem
\begin{equation}
\hat{A}\Phi=\lambda\Phi,
\end{equation}
where $\hat{A}$ is a operator of the form
\begin{equation}
\hat{A}=a_{0}K_{0}+a_{1}K_{+}+a_{2}K_{-}.\label{AHsu11}
\end{equation}
In order to do this, we apply a similarity transformation to this equation in terms of the displacement operator (\ref{do}) to obtain
\begin{equation}
\hat{A}'\Phi'=\lambda\Phi',
\end{equation}
where the eigenvalue $\lambda$ remains unchanged, the eigenfunction is transformed as $\Phi'=D^{\dag}(\xi)\Phi$ and the operator $\hat{A}$ becomes
\begin{align}
\hat{A}'=D^{\dag}(\xi)\hat{A}D(\xi)&=\left[(2\beta+1)a_{0}+\frac{\xi^{*}}{|\xi|}\alpha a_{1}+\frac{\xi}{|\xi|}\alpha a_{2}\right]K_{0}\nonumber\\&+\left[\frac{\xi}{2|\xi|}\alpha a_{0}+(\beta+1)a_{1}+\frac{\xi}{\xi^{*}}\beta a_{2}\right]K_{+}\nonumber\\&+\left[\frac{\xi^{*}}{2|\xi|}\alpha a_{0}+\frac{\xi^{*}}{\xi}\beta a_{1}+(\beta+1)a_{2}\right]K_{-},\label{A'}
\end{align}
with $\alpha=\sinh(2|\xi|)$ and $\beta=\frac{1}{2}\left[\cosh(2|\xi|)-1\right]$.
Then, by choosing the coherent parameters $\theta$ and $\phi$ of the complex number $\xi=-\frac{\theta}{2}e^{-i\phi}$ as
\begin{equation}
\theta=\tanh^{-1}\left[\frac{2\sqrt{a_{1}a_{2}}}{a_{0}}\right],\quad\quad\phi=i\ln\left[\frac{a_{o}\alpha}{2a_{2}(2\beta+1)}\right] ,
\end{equation}
we can remove the operators $K_{\pm}$ in the equation (\ref{A'}). Hence, the eigenvalue equation is reduced to
\begin{equation}
\sqrt{a_{0}^{2}-4a_{1}a_{2}}K_{0}\Phi'=\lambda\Phi'.\label{ADHsu11}
\end{equation}
Finally, if $\Phi'$ is a eigenfunction of the operator $K_{0}$, we have solved the eigenvalue problem. Therefore, it is convenient that $K_{0}$ is an operator such that we know its eigenfunctions and eigenvalues. Moreover, notice that if the coefficients $a_{0}$, $a_{1}$ and $a_{2}$ are equal, the problem does not have a exact solution.

\renewcommand{\theequation}{B.\arabic{equation}}
\setcounter{equation}{0}

\section*{Appendix B. The $SU(2)$ group theory and its tilting transformation}

The $su(2)$ Lie algebra is spanned by the generators $J_{+}=J_1+iJ_2$, $J_{-}=J_1-iJ_2$ and $J_{0}$, which satisfy the commutation relations \cite{Vourdas}
\begin{eqnarray}
[J_{0},J_{\pm}]=\pm J_{\pm},\quad\quad [J_{+},J_{-}]=2J_{0}.\label{com2}
\end{eqnarray}
The displacement operator $D(\xi)$ for this Lie algebra is defined as
\begin{equation}
D(\xi)=\exp(\xi J_{+}-\xi^{*}J_{-}),\label{D}
\end{equation}
where $\xi=-\frac{1}{2}\theta e^{-i\varphi}$. In this case, the normal form of this operator is given by
\begin{equation}
D(\xi)=\exp(\zeta J_{+})\exp(\eta J_{0})\exp(-\zeta^*J_{-})\label{normal2},
\end{equation}
where  $\zeta=-\tanh(\frac{1}{2}\theta)e^{-i\varphi}$ and $\eta=-2\ln \cosh|\xi|=\ln(1-|\zeta|^2)$. The $SU(2)$ Perelomov coherent states $|\zeta\rangle=D(\xi)|j,-j\rangle$ are defined in terms of  $D(\xi)$ as \cite{PERL,Arecchi}
\begin{equation}
|\zeta\rangle=\sum_{\mu=-j}^{j}\sqrt{\frac{(2j)!}{(j+\mu)!(j-\mu)!}}(1+|\zeta|^{2})^{-j}\zeta^{j+\mu}|j,\mu\rangle.\label{PCN2}
\end{equation}
The $SU(2)$ Perelomov number coherent state $|\zeta,j,\mu\rangle$ is defined as the action of the displacement operator $D(\xi)$ on an arbitrary
excited state $|j,\mu\rangle$ \cite{Nos3}
\begin{eqnarray}
|\zeta,j,\mu\rangle &=&\sum_{s=0}^{j-\mu+n}\frac{\zeta^{s}}{s!}\sum_{n=0}^{\mu+j}\frac{(-\zeta^*)^{n}}{n!}e^{\eta(\mu-n)}
\frac{\Gamma(j-\mu+n+1)}{\Gamma(j+\mu-n+1)}\nonumber\\ &&\times\left[\frac{\Gamma(j+\mu+1)\Gamma(j+\mu-n+s+1)}{\Gamma(j-\mu+1)\Gamma(j-\mu+n-s+1)}\right]^{\frac{1}{2}}|j,\mu-n+s\rangle.\label{PNCS2}
\end{eqnarray}

The tilting transformation of the $su(2)$ Lie algebra generators are computed by using the displacement operator $D(\xi)$ to obtain
\begin{equation}
D^{\dag}(\xi)J_{+}D(\xi)=-\frac{\xi^{*}}{|\xi|}\delta J_{0}+\epsilon\left(J_{+}+\frac{\xi^{*}}{\xi}J_{-}\right)+J_{+},
\end{equation}
\begin{equation}
D^{\dag}(\xi)J_{-}D(\xi)=-\frac{\xi}{|\xi|}\delta J_{0}+\epsilon\left(J_{-}+\frac{\xi}{\xi^{*}}J_{+}\right)+J_{-},
\end{equation}
\begin{equation}
D^{\dag}(\xi)J_{0}D(\xi)=(2\epsilon+1)J_{0}+\frac{\delta\xi}{2|\xi|}J_{+}+\frac{\delta\xi^{*}}{2|\xi|}J_{-},
\end{equation}
where $\delta=\sin(2|\xi|)$ and $\epsilon=\frac{1}{2}\left[\cos(2|\xi|)-1\right]$.

The Jordan-Schwinger realization of the $su(2)$ algebra is given by the operators
\begin{equation}
J_0=\frac{1}{2}\left(a^{\dag}a-b^{\dag}b\right), \quad J_+=a^{\dag}b, \quad J_-=b^{\dag}a,\label{gen}
\end{equation}
where again, the two sets of operators $(a, a^{\dag})$ and $(b, b^{\dag})$ satisfy the bosonic algebra.
It is important to note that, besides the Casimir operator, there is another operator $N_s$ (called the number operator)
which commutes with all the generators of the $su(2)$ algebra.

The Casimir operator $J^2$ for this realization is written in terms of $N_s$ and the two operators explicitly are
\begin{equation}
J^2=\frac{N_s}{2}\left(\frac{N_s}{2}+1\right),\quad\quad N_s=a^{\dag}a+b^{\dag}b,\nonumber
\end{equation}
\begin{equation}
[N_s,J_+]=[N_s,J_-]=[N_s,J_z]=0.
\end{equation}

Analogously to the theory presented in Appendix A, if $\hat{A}$ is a operator of the form
\begin{equation}
\hat{A}=a_{0}J_{0}+a_{1}J_{+}+a_{2}J_{-}, \label{AHsu2}
\end{equation}
the $SU(2)$ displacement operator can be used to solve the eigenvalue problem
\begin{equation}
\hat{A}\Phi=\lambda\Phi.\label{Asu2}
\end{equation}
By applying the displacement operator (\ref{D}) to the eigenvalue equation, we have that it becomes
\begin{equation}
\hat{A}'\Phi'=\lambda\Phi',
\end{equation}
where $\Phi'=D^{\dag}(\xi)\Phi$ and the operator $\hat{A}$ transforms as
\begin{align}
\hat{A}'=D^{\dag}(\xi)\hat{A}D(\xi)&=\left[(2\epsilon+1)a_{0}-\frac{\xi^{*}}{|\xi|}\delta a_{1}-\frac{\xi}{|\xi|}\delta a_{2}\right]J_{0}\nonumber\\&+\left[\frac{\xi}{2|\xi|}\delta a_{0}+(\epsilon+1)a_{1}+\frac{\xi}{\xi^{*}}\epsilon a_{2}\right]J_{+}\nonumber\\&+\left[\frac{\xi^{*}}{2|\xi|}\delta a_{0}+\frac{\xi^{*}}{\xi}\epsilon a_{1}+(\epsilon+1)a_{2}\right]J_{-},
\end{align}
with $\delta=\sin(2|\xi|)$ and $\epsilon=\frac{1}{2}\left[\cos(2|\xi|)-1\right]$.
Then, we can remove the operators $J_{\pm}$ by choosing the parameters $\theta$ and $\phi$ of the complex number $\xi=-\frac{\theta}{2}e^{-i\phi}$ as
\begin{equation}
\theta=\arctan\left(\frac{2\sqrt{a_{1}a_{2}}}{a_{0}}\right),\quad\quad\phi=i\ln\left[\frac{a_{o}\delta}{2a_{2}(2\epsilon+1)}\right].
\end{equation}
With this particular choice the eigenvalue equation is reduced to
\begin{equation}
\sqrt{a_{0}^{2}+4a_{1}a_{2}}J_{0}\Phi'=\lambda\Phi'. \label{ADHsu2}
\end{equation}
Finally, if $\Phi'$ is a eigenfunction of the operator $J_{0}$, we have solved the eigenvalue problem of equation (\ref{Asu2}).

\end{document}